\RequirePackage{lineno}

\documentclass[%
 reprint,
groupedaddress,
 amsmath,amssymb,
 aps,
]{revtex4-2}

\usepackage{graphicx}
\usepackage{dcolumn}
\usepackage{bm}
\usepackage{ulem}
\usepackage{xfrac}
\usepackage{color}
\usepackage{graphicx,amsmath,amssymb}
\usepackage{epstopdf}
\usepackage{lineno}
\usepackage[colorlinks=true, urlcolor=blue, linkcolor=blue, citecolor=blue]{hyperref}

\definecolor{turquoiseblue}{rgb}{0.02, 0.55, 0.55}
\definecolor{green}{rgb}{0.02, 0.55, 0.55}

\begin{document}

\title{
New Analysis of Dark Matter in Elliptical Galaxies}
\author{D.~M.~Winters$^1$, A.~Deur$^{1,2}$, X. Zheng$^1$\\
\vspace{5pt}
    $^1$University of Virginia, Charlottesville, Virginia 22904, USA \\
    $^2$Old Dominion University, Norfolk, VA 23508, USA\\
}

\begin{abstract}

 We investigate a correlation between the dark matter content of elliptical galaxies and their ellipticity $\epsilon$ that was initially reported in 2014. We use new determinations of dark matter and ellipticities that are posterior to that time. Our data set consists of 237 elliptical galaxies passing a strict set of criteria. We find a relation 
 between the mass-to-light ratio and ellipticity $\epsilon$ that is well fit by $\sfrac{M}{L}=(14.1\pm 5.4)\epsilon$, which agrees with the result reported in 2014.\\
 \textbf{Key words:} galaxies: structure - cosmology: dark matter

\end{abstract}


\maketitle

\section{Introduction}

Dark matter is an essential element of galaxy dynamics. The understanding that galaxies 
are constituted of a large halo of dark matter encompassing a much smaller baryonic visible component originated from disk galaxies. 
There, the organized motion of stars and gas allows us to deduce 
the total galactic mass distribution. Assessing the dark matter content of an elliptical galaxy is more challenging because
the orbital plane of its stars are random and our observations typically do not resolve stars 
in galaxies other than the Milky-way and its satellites. 
However, methods have been devised to measure the total mass of elliptical galaxies such as Jeans-Anisotropic Modeling~\cite{pechetti2017detection, poci2017systematic, shetty2020precise, ouellette2017spectroscopy}, or through observations of gravitational lensing~\cite{shu2017sloan, leier2016strong, sonnenfeld2015sl2s}, globular clusters~\cite{alabi2017sluggs, harris2019globular}, X-ray emissions~\cite{jin2020sdss}, or companion galaxies~\cite{chen2020dynamics}. 
Intriguingly, it was reported that some elliptical galaxies have little dark mass~\cite{romanowsky2003dearth}
 and that the dark mass of elliptical galaxies appears correlated with the galaxy ellipticity $\epsilon$~\cite{Deur:2013baa}: the rounder the
galaxy, the less dark matter it seems to contain. These findings are puzzling:
A low dark matter content of round galaxies contradicts 
the mechanism of galaxy formation, which demands that the primordial gas 
aggregates in a dark matter halo that had collapsed at earlier times. This is compounded by the scenario that many elliptical galaxies originate
from galactic mergers. 
Furthermore, one expects halos to be more or less spherical  
and to dominate galaxy dynamics, regardless of the galaxy types. There is then no obvious mechanism to correlate the visible shape of a galaxy to the relative mass of its spherical dark halo.  
In this article, we re-examine the dark mass-ellipticity correlation found in~\cite{Deur:2013baa} using determinations of dark matter content published after that study.

In Ref.~\cite{Deur:2013baa}, a large and homogeneous data set (685 galaxies) was used that excluded galaxies showing unusual features, 
e.g., an Active Galactic Nucleus (AGN). This was to ensure that the study 
would not be biased by unusual or disturbed elliptical galaxies or by S0 galaxies. 
The analysis used dark masses assessed from the methods available at the time, namely (1) virial analyses, (2) orbit modeling, and observations of (3) globular clusters, (4) X-ray emissions, (5) embedded disk dynamics and (6) lensing. 
Remarkably, the results based on each
methods displayed a positive correlation, albeit
the uncertainties on the results from methods (2) and (5) are large. 
This indicates that methodological biases are not likely to have generated the
correlation. Finally, it was checked that the correlation did not arise from the propagation of other physical correlations, 
e.g., a first correlation between the dark mass and characteristic $A$ of the galaxy and a second one between $A$ and $\epsilon$. 
Such check 
suggests a physical origin for the correlation, which would demand a revision of 
the structure of the
galactic dark halo, of the dark matter paradigm, or of the galaxy formation and evolution scenarios. 
The alternative conclusion, that there is a significant methodological bias common to all the methods used to 
assess the dark mass of elliptical galaxies, would also have important consequences. This would mean that biased methods are used to analyse galactic structure and evolution. 
Consequently, it is crucial to verify the existence of the dark mass-$\epsilon$ correlation. 
New data, with improved observations and methods, have become available since 
the analysis reported in~\cite{Deur:2013baa}.  They will shed new light upon the dark mass-$\epsilon$ correlation, the object of this article.

\section{Method \label{sec:method}}

We follow the same method as in~\cite{Deur:2013baa}, recapitulated here for convenience. 
We use 17 separate publications that study at least several elliptical galaxies and provide their 
dark matter fraction ($DMf$), mass-to-light $\sfrac{M}{L}$ or total mass-to-stellar mass $\sfrac{M}{M_*}$ ratios. 
We will refer globally to these quantities as the {\it dark matter content} (DMC).
After rejecting galaxies according to criteria summarized below, 
we obtained 237 different bona-fide galaxies (or 329 galaxies given the same galaxy may appear in different articles), compared to the 255 (or 685) galaxies considered in~\cite{Deur:2013baa}.

For each publication, we select a homogeneous sample of elliptical galaxies by choosing those without outstanding features. 
We remove those with peculiar attributes such as AGN to suppress noise from atypical or dynamically disturbed galaxies. Noise in the data set could either drown out any existing correlation or cause an artificial one. 
Different selection criteria are applied depending on whether galaxies are local or distant. Local
galaxies are typically better characterized, which allows for stricter selection criteria. Conversely, selection criteria of distant galaxies
(typically used in gravitational lensing studies) are less strict. 
The galaxy characteristics used for selection are obtained from the 
NASA/IPAC Extragalactic Database (NED)~\cite{ipac} or from the publication that analyzed the galaxy DMC.
Local galaxies are selected to be medium size elliptical galaxies without sign of disturbance.
 Galaxies not considered in this analysis include:
 
1. S0 galaxies, compact elliptical galaxies (cE), giant elliptical galaxies (D, cD) and Brightest Cluster Galaxies (BrClG),
transition-type (E+), E? galaxies as listed in NED. If NED lists several modern ($\gtrsim1970$) 
publications with conflicting galactic class, the galaxy is removed from the analysis;

2.  galaxies with AGN, HII emission lines or NELG (Narrow Emission Line Galaxy);

3.  LINER (Low-Ionization Nuclear Emission-line Region), Seyfert (Sy) and BL Lacertae (BLLAC) galaxies;

4. peculiar galaxies with additional reference from the Arp catalogue~\cite{Arp}.\\
However, LERG (Low Excitation Radio Galaxy) and WLRG (Weak emission-Line Radio-Galaxy) are kept. 

In contrast, distant galaxies are typically not characterized well enough to be subjected to
the criteria listed above. 
Yet, they are important for the present study since their
large number enables the 
use of gravitational lensing to determine galactic total masses. The following distant galaxies are not considered in this analysis:

1.  massive galaxies ($M_*\gtrsim5\times10^{11}\mbox{M}_{\odot}$), to reject cD, D or BrClG galaxies; 

2. galaxies with velocity dispersions $\sigma \leq225$~km \, s$^{-1}$, to reject possibly S0 galaxies.\\
Furthermore, if an attribute used for rejection of local galaxies, e.g., if it has an AGN,
is available from NED or the publication analyzing the galaxy mass, then that distant galaxy is also not considered. 

The effect of these criteria is that a large fraction $\geq$ 90\%
of the galaxies included in the articles are not considered in our analysis.

For each article, we collect the ellipticities of the galaxies from the values provided in the article, mass ellipticities from Bolton et al.~\cite{Bolton:2008xf} (lensing galaxies), or ellipticities given from NED (in order of preference). We collect the DMC from the articles as $\sfrac{M}{L}$, $DMf$, or $M/M_* = 1/(1-DMf)$ and when different DMC types are provided, we choose $\sfrac{M}{L}$. 
Finally, we assess the uncertainties on the DMC in one of three ways.
If the DMC are reported with uncertainties, and that a linear fit of these DMC vs $\epsilon$ returns a $\chi^2 / ndf < 1$, we homogeneously scale the uncertainties until $\chi^2 / ndf = 1$, according to the {\it unbiased estimate} procedure~\cite{Schmelling:1994pz, particle2020review}. 
The remaining two cases occur when no uncertainties are provided. In the case of $M/L$, we add an uncertainty proportional to $M/L$ until $\chi^2 / ndf = 1$~\footnote{In addition,
Ref.~\cite{poci2017systematic} is specially treated because it provides the $DMf$ up to one effective radius only, thereby obtaining much smaller $DMf$ values than in the other articles. Thus for~\cite{poci2017systematic}, we add a linear uncertainty, $P*\epsilon + 0.1$, where we scale $P$ until $\chi^2/ndf=1$, rather than an uncertainty proportional to the $DMf$. 
}. In the case of the $DMf$, since its value can be zero, 
we add a constant contribution so that the $DMf$ vs $\epsilon$ fit is not almost exclusively driven by points of the smallest $DMf$. This additional contribution to the uncertainty is again determined by requiring $\chi^2 / ndf = 1$.
The {\it unbiased estimate} is a convenient and simple method to evaluate uncertainties not reported in an article or combining uncertainties of unknown reliability or correlations. It can be straightforwardly applied to all data sets for analysis consistency. It is in our case a conservative procedure since the 
uncertainties provided by the publications always needed to be increased to reach $\sfrac{\chi^{2}}{ndf}=1$. Finally, its simplicity
minimizes the chance of introducing biases.

To quantify the correlation between DMC and $\epsilon$, we
linearly fit the DMC plotted versus $\epsilon$. A clear non-zero slope of the fit then reveals a correlation.
A linear form for the fit was found to be sufficient in Ref.~\cite{Deur:2013baa}. The fit is performed on each data set by linear regression, weighted by uncertainties. We then record both the slope and its uncertainty for each data set. 
To average these slope values requires care. Different DMC analyses may yield different results for the same galaxy for multiple reasons:

    $\bullet$ different choices of band for the galactic luminosity;
    
    $\bullet$ using $\sfrac{M}{L}$, $M/M_*$ or $DMf$;
    
    $\bullet$ using different maximum radii to integrate the DMC;
    
    $\bullet$ using different Hubble parameter values;
    
    $\bullet$ possible systematic biases in analysis methods.\\
For instance, an article that provides DMC out to one effective radius will obtain much lower DMC values than another article measuring out to five effective radii. To account for such variations, we normalize each article's DMC to a given value. We treat $\sfrac{M}{L}$ and $M/M_*$ identically, but $DMf$ must be transformed into $M/M_*$ before it can be normalized. As in \cite{Deur:2013baa}, we homogeneously scale the DMC of an article until the linear regression satisfies $\sfrac{M}{L}|_{\epsilon  = 0.3}=8~\sfrac{M_\odot}{L_\odot}$ and $M/M_*|_{\epsilon=0.3}=8$.
The procedure assumes that the systematic differences
instanced above are $\epsilon$-independent.
We also remark that each slope $d(\sfrac{M}{L})/d\epsilon$ scales with the normalization value.

Another necessary correction is to account for the projection of a galaxy onto our field of view. 
An observed galaxy tends to appear rounder because the observed ellipticity is a projection of the actual ellipticity. This reduces the DMC-$\epsilon$ correlation. We account for it by scaling the slope values by a projection correction, derived in~\cite{Deur:2013baa}, which found that the projection reduces the slope $d(\sfrac{M}{L})/d\epsilon$ of the fit by a factor of $5\pm1$.
 
One further correction is that different data sets may have common galaxies. We account for such overlap by assigning to each galaxy a weight $N_i$, defined as the number of data sets using the $i^{\rm th}$ galaxy. The slope uncertainty is then scaled by $\sqrt{N_p / [\Sigma_{i \in p} (1/ N_i) ] }$ where $N_p$ is the number of galaxies for data set $p$.

To average properly the $\sfrac{M}{L}(\epsilon)$ slopes among data sets, the DMC uncertainties must be estimated consistently, lest articles with conservative slope uncertainties be unwarrantedly weighted out. We again apply the {\it unbiased estimate} on the slope uncertainties after the normalization of DMC to 8 $\sfrac{M_\odot}{L_\odot}$, the corrections for projection effect and adjustment for shared galaxies.
This can be done by 
either globally rescaling the uncertainties or adding a constant uncertainty to each slope error.
The {\it unbiased estimate} applies ideally when uncertainties cause
a Gaussian dispersion of the data around the expectation. 
Due to systematic effects (such as the random projection of the actual 3D galactic ellipsoid into the observed 2D ellipse, which turns the presumed Gaussian distribution of true ellipticity into a non-Gaussian distribution favoring small apparent ellipticities.) the data scatter is unlikely to be truly Gaussian.
However, numerous independent systematic effects of similar size will result in an approximate Gaussian distribution. 

\section{Description of the data sets} 

We use data sets that were produced after the analysis of Ref.~\cite{Deur:2013baa} in 2013. 
The data sets are published in
17 articles~\cite{pechetti2017detection, poci2017systematic, shetty2020precise, ouellette2017spectroscopy, shu2017sloan, leier2016strong, sonnenfeld2015sl2s, alabi2017sluggs, harris2019globular, jin2020sdss, chen2020dynamics, sanders2014dearth, aquino2018kinematic, belli2016mosfire, sonnenfeld2019survey, tian2017mass, dabringhausen2016extensive}, each with at least 3 galaxies fulfilling the criteria described in Section~\ref{sec:method}. Each article uses one of five different methods for measuring galactic DMC:

1.~The most common method is Penalized-Pixel Fitting (PPXF) to extract stellar kinematics and then using Jeans Anisotropic Modeling (JAM) to determine the total mass of a galaxy~\cite{ppxf,jam}. This was employed in 6 articles~\cite{pechetti2017detection, poci2017systematic, shetty2020precise, aquino2018kinematic, belli2016mosfire, ouellette2017spectroscopy}. 

2.~The next most common method is gravitational lensing, used in 5 articles~\cite{sanders2014dearth, shu2017sloan, leier2016strong, sonnenfeld2015sl2s, sonnenfeld2019survey, tian2017mass}.

3.~Two articles~\cite{alabi2017sluggs, harris2019globular} determined DMC from the observation of globular cluster motions. 

4.~Ref.~\cite{chen2020dynamics} used the motion of companion galaxies. 

5.~Ref.~\cite{jin2020sdss} primarily used galactic X-ray data and partially checked against a globular cluster method.\\
We also include Ref.~\cite{dabringhausen2016extensive} that collects data from several surveys and thus calculates DMC using a mixture of methods. We remark that some of the methods used in articles employed by Ref.~\cite{Deur:2013baa} are not used in the more recent articles, namely the method using the virial theorem, that using planetary nebulae and that using gas disk dynamics.

\begin{figure*}[!htp]
\centering
\centerline{\includegraphics[scale=0.45]{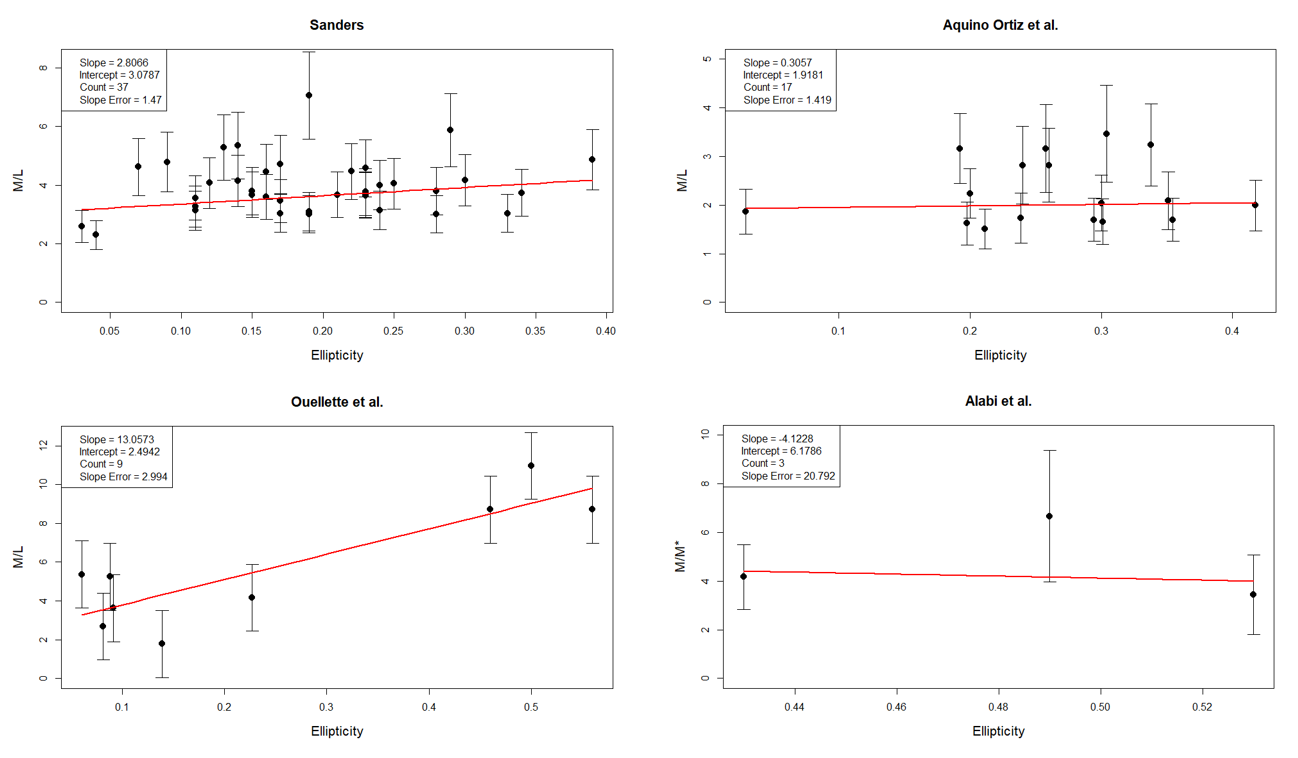}}
\caption{\label{fig: MLplots}
Examples of Mass-to-Light ratio (in units of $\sfrac{M_\odot}{L_\odot}$) or $M/M_*$ vs ellipticity, with the $\sfrac{M}{L}$ determined in Refs.~\cite{sanders2014dearth, aquino2018kinematic,ouellette2017spectroscopy,alabi2017sluggs}. The red line shows the best linear fit to the data with its slope and intercept given in the top left insert. }
\end{figure*}

As stated previously, the selection criteria remove the vast majority of galaxies analyzed in the articles. 
Some post-2013 articles were not included in our analysis because they were superseded by more recent articles (same data set and method), Ref.~\cite{kim2019revisiting, forbes2016sluggs}, or because there remained fewer than three galaxies after selection~\cite{bilek2019study}. 
Figure~\ref{fig: MLplots} displays examples of $\sfrac{M}{L}$ vs $\epsilon$ plots before they are normalized in order to combine them together. The other plots not shown in Fig.~\ref{fig: MLplots} can be found in the Appendix.

\section{Global results \label{sec:Global-results}}
\begin{figure*}[!htp]
\centering
\centerline{\includegraphics[scale=0.45]{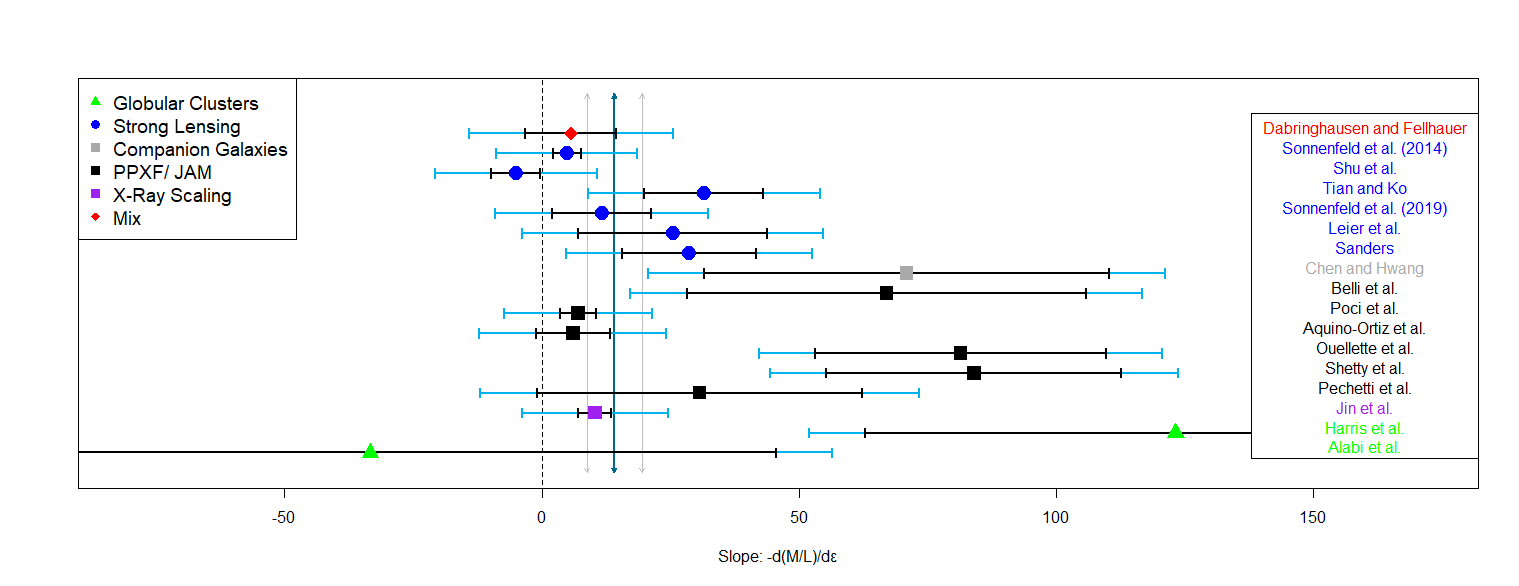}}
\caption{\label{fig: global}The $d(\sfrac{M}{L})/d\epsilon$ slopes  ($\sfrac{M_\odot}{L_\odot}$ unit)
for the 17 data sets after projection correction and normalization
to a common $\sfrac{M}{L}|_{\epsilon =0.3}$ value.
The different symbols denote the different methods used to extract the DMC of the
galaxies.
The inner (black) and outer (blue) error bars are, respectively, the uncertainties before and after requiring that
the $d(\sfrac{M}{L})/d\epsilon$ dispersion is Gaussian ({\it unbiased estimate} procedure).
The vertical continuous lines indicate the average value $\langle \sfrac{d(\sfrac{M}{L})}{d\epsilon}\rangle $ and its uncertainty, with averaging done using the full uncertainty (blue error bars). The vertical dash line indicates zero, for reference.
}
\end{figure*}

The fit results on individual data sets are 
normalized to 
$\sfrac{M}{L} |_{\epsilon =0.3} = 8 \sfrac{M_\odot}{L_\odot}$, $ \sfrac{M}{M_*} |_{\epsilon =0.3}=8$, and then
corrected for the $\epsilon$-projection effect.
The uncertainties on the individual fit slopes $\sfrac{d(\sfrac{M}{L})}{d\epsilon}$ are increased to
account for the common galaxies shared by different data sets. 
%
This provides the $\sfrac{d(\sfrac{M}{L})}{d\epsilon}$ shown in Fig.~\ref{fig: global} with the inner (black) error bars.
We apply the {\it unbiased estimate} on the data in Fig.~\ref{fig: global} where we added a constant contribution to the black error bars rather than scaling them so that small DMC values do not dominate the global fit. 
 The final uncertainties are shown by the outer blue error bars in Fig.~\ref{fig: global}.
This last application of the {\it unbiased estimate} adds an uncertainty contribution for possible systematic bias in each of the publications,
and corrects for any possible uncertainty overestimate when performing global corrections (e.g.,
projection corrections) or when combining uncertainties. 

The individual $\sfrac{d(\sfrac{M}{L})}{d\epsilon}$ are then
combined to form the average slope of $\sfrac{M}{L}(\epsilon)$, yielding $\langle \sfrac{d(\sfrac{M}{L})}{d\epsilon}\rangle = (14.1\, \pm\, 5.4)\sfrac{M_\odot}{L_\odot}$, a statistically meaningful correlation between $\sfrac{M}{L}$ and $\epsilon$. 

It is useful to know the effects of the various corrections applied in this analysis. 
The average slope without projection correction, correction for shared
galaxies or applying the {\it unbiased estimate} on the $\sfrac{d(\sfrac{M}{L})}{d\epsilon}$ data is
$\langle \sfrac{d(\sfrac{M}{L})}{d(\epsilon)}\rangle = (2.67 \pm 0.61)\sfrac{M_\odot}{L_\odot}$, i.e., a non-zero slope with 4.4$\sigma$
significance~\footnote{The significance assumes a Gaussian meaning to the uncertainty. It is reasonable since the uncertainties reported in the publications for each data set are largely independent and furthermore a significant
Gaussian uncertainty is added to them when applying the {\it unbiased estimate}. 
In any case, $\sigma$ is used here simply to compare the significance of the slopes after the various corrections are applied. Therefore, the assumption
whether the uncertainties are dominantly Gaussian or not is not crucial.}.
The result with the projection correction but without shared
galaxies or {\it unbiased estimate} corrections on the $\sfrac{d(\sfrac{M}{L})}{d\epsilon}$ is
 $\langle \sfrac{d(\sfrac{M}{L})}{d(\epsilon)}\rangle = (8.5 \pm 1.4)\sfrac{M_\odot}{L_\odot}$ (6.1 $\sigma$ significance).
The result with projection correction and applying the {\it unbiased estimate} on the $\sfrac{d(\sfrac{M}{L})}{d\epsilon}$, but no correction 
for shared galaxies is $(17.1 \pm 5.4)\sfrac{M_\odot}{L_\odot}$ (3.2 $\sigma$ significance).
When we also account for shared galaxies we obtain $(14.1\, \pm\,  5.4)\sfrac{M_\odot}{L_\odot}$ (2.6 $\sigma$ significance), our final average slope value. This shows that overall, the corrections have not artificially increased the significance of the correlation but
decreased it due to our conservative estimates when determining uncertainties, particularly that of the {\it unbiased estimate}.

Figure~\ref{fig: global} emphasizes that individual methods for measuring DMC have a positive $\langle d(\sfrac{M}{L})/d\epsilon \rangle$. In particular the two most used methods, lensing and PPXF/JAM yield $(9.5 \, \pm \, 6.4)\; \sfrac{M_\odot}{L_\odot}$ and $(22.9 \pm 9.0)\;\sfrac{M_\odot}{L_\odot}$, respectively.
 Our slope of $(14.1\, \pm\, 5.4) \;\sfrac{M_\odot}{L_\odot}$ agrees with the result $(14.5\pm 4.8) \;\sfrac{M_\odot}{L_\odot}$  from~\cite{Deur:2013baa}. 
The near match of the two results despite their relatively large uncertainties could be a coincidence, but it could also be that using methods that yield conservative uncertainties overestimates our uncertainty and that of~\cite{Deur:2013baa}. For example, by accounting for overlap of galaxies and adding a constant uncertainty (in addition to shifting the global $\langle d(\sfrac{M}{L})/d\epsilon \rangle$ average), the uncertainty on the mean increases by a factor of 4.

\section{Stellar $\sfrac{M_*}{L}$ ratio vs ellipticity}

\begin{figure}[h]
\centering
\includegraphics[scale=0.35]{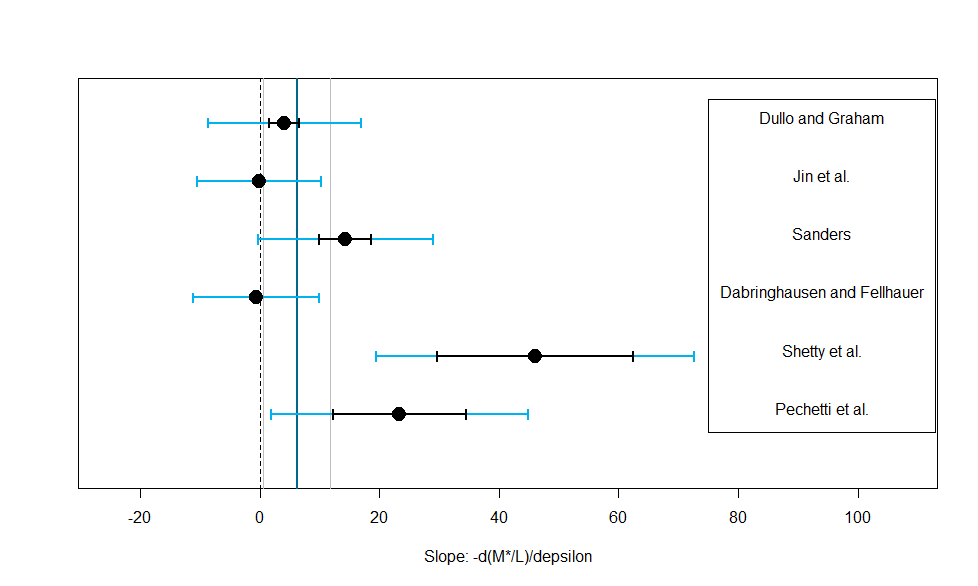}
\caption{\label{fig: global-stelar}
Same as Fig.~\ref{fig: global} but for the stellar $\sfrac{d(\sfrac{M_*}{L})}{d\epsilon}$.}
\end{figure}

Although the possibility of a correlation between galactic missing mass and ellipticity had been suggested before~\cite{Deur:2009ya}, such correlation is unexpected in the standard scenario of galaxy formation. 
Thus, for a positive finding of the correlation to be credible, one must strive to avoid systematic biases and 
conduct systematic studies. Ref.~\cite{Deur:2013baa} performed numerous systematic checks but it did not include an investigation  
of a possible $\sfrac{M_*}{L}$-$\epsilon$ correlation. Since the stellar mass $M_*$ is obtained from $L$, 
no correlation should be found, lest methods or data are biased, e.g., with 
S0 contamination~\footnote{Like $\sfrac{M}{L}$, $\sfrac{M_*}{L}$
is usually smaller for S0 than for elliptical galaxies. Ref.~\cite{Cappellari:2012ad} reported
that early-type galaxies with low velocity dispersion, {\it viz}, those
more likely to be S0, have $\sfrac{M_*}{L} \simeq 2 \sfrac{M_{\odot}}{L_{\odot}}$, while 
those with high velocity dispersion, {\it viz}, more likely to be ellipticals, 
have $\sfrac{M_*}{L} \simeq 4 \sfrac{M_{\odot}}{L_{\odot}}$.%
Thus, if S0 contamination generated the $\sfrac{M}{L}$-$\epsilon$
 correlation, a $\sfrac{M_*}{L}$-$\epsilon$ correlation would also be observed.}.
For such check, we use the publications employed in~\cite{Deur:2013baa} and in the present study that provide
$\sfrac{M_*}{L}$ in addition to DMC. We first discuss the check for~\cite{Deur:2013baa}  (see Ref.~\cite{deur2020correlation} for details) and then for the present study.\\
$\mathbf{\sfrac{M_*}{L}(\epsilon)}$ \textbf{study for Ref.~\cite{Deur:2013baa}'s analysis}
Of the 41 data sets used in Ref.~\cite{Deur:2013baa}, 13 of them also provide $\sfrac{M_*}{L}$:
Auger {\it et al.}~\cite{Auger1},
Barnabe {\it et al.}~\cite{Barnabe},
Capaccioli {\it et al.}~\cite{Capaccioli},
Cappellari {\it et al.} (2006)~\cite{Cappellari06},
Cappellari {\it et al.} (2013)~\cite{Cappellari13a},
Cardone {\it et al.}~\cite{Cardone09},
Conroy and van Dokkum~\cite{Conroy},
Deason {\it et al.}~\cite{Deason},
Grillo {\it et al.}~\cite{Grillo09},
Jiang and Kochanek~\cite{Jiang-Kochanek},
Leier {\it et al.}~\cite{Leier2011},
Thomas {\it et al.}~\cite{Thomas} together with Wegner {\it et al.} (2012)~\cite{Wegner},
Treu and Koopman~\cite{Treu04}.
Uncertainties are taken to be proportional~\footnote{
Except for Ref.~\cite{Cappellari:2012ad} for which we use a constant 
uncertainty so that the fit is not dominated
by small $\Delta \sfrac{M_*}{L}$.}  
to $\sfrac{M_*}{L}$ and scaled according to the {\it unbiased estimate}.
Before combining the individual $d(\sfrac{M_*}{L})/d\epsilon$, 
we normalize $\sfrac{M_*}{L} |_{\epsilon = 0.3}$ to 4 $\sfrac{M_{\odot}}{L_{\odot}}$, about the expected average value in the B-band.
The resulting average $\langle \sfrac{d(\sfrac{M_*}{L})}{d\epsilon} \rangle = (-0.14\pm1.19)\sfrac{M_\odot}{L_\odot}$ is compatible with zero. \\
$\mathbf{\sfrac{M_*}{L}(\epsilon)}$ {\bf study for the present analysis} For the present study, 5 out of 17 data sets also include $\sfrac{M_*}{L}$: 
Dabringhausen and Fellhauer~\cite{dabringhausen2016extensive}, 
Jin {\it et al.}~\cite{jin2020sdss},
Pechetti {\it et al.}~\cite{pechetti2017detection},
Sanders~\cite{Sanders:2013pfa} and
Shetty {\it et al.}~\cite{shetty2020precise}. 
We also use Dullo and Graham~\cite{Dullo:2013hfa} that provides $\sfrac{M_*}{L}$ but no DMC.
We perform the same analysis as for $\sfrac{M}{L}$ on all 6 data sets except that we normalize to $\sfrac{M_*}{L}|_{\epsilon =0.3} = 4$ instead of 8.
We find $\langle \sfrac{d(\sfrac{M_*}{L})}{d\epsilon} \rangle =(6.2 \pm 5.5)\;\sfrac{M_\odot}{L_\odot}$ which is not a clear sign of correlation, especially with only 6 data points. We also note that before applying the {\it unbiased estimated}, the value is $-0.054 \pm 0.075$ because Refs.~\cite{dabringhausen2016extensive} and \cite{jin2020sdss} have small error bars. In Fig.~\ref{fig: global-stelar} these are represented by the black set of error bars.

The two $\sfrac{d(\sfrac{M_*}{L})}{d\epsilon}$ values $(-0.14\pm1.19)\sfrac{M_\odot}{L_\odot}$ from the pre-2013 publications~\cite{Auger1, Barnabe, Capaccioli, Cappellari06, Cappellari13a, Cardone09, Conroy, Deason, Grillo09, Jiang-Kochanek, Leier2011, Thomas, Wegner, Treu04} and 
$(6.2 \pm 5.5)\sfrac{M_\odot}{L_\odot}$ from the post-2013 publications~\cite{dabringhausen2016extensive, jin2020sdss, pechetti2017detection, Sanders:2013pfa, shetty2020precise, Dullo:2013hfa} average to 
$\sfrac{d(\sfrac{M_*}{L})}{d\epsilon}=(0.14 \pm 1.16)\sfrac{M_\odot}{L_\odot}$, a slope compatible
with 0 and two orders of magnitude smaller than $\sfrac{d(\sfrac{M}{L})}{d\epsilon}=14.3\pm 3.6$, the average result from Ref.~\cite{Deur:2013baa} and the 
present analysis. If the cause of
the $\sfrac{M}{L}(\epsilon)$ correlation affected equaly $\sfrac{M_*}{L}$, then given the $\sfrac{M_*}{L}|_{\epsilon =0.3}=0.5 \sfrac{M}{L}(0.3$) normalizations, one would have had $\sfrac{d(\sfrac{M_*}{L})}{d\epsilon} \simeq 0.5\sfrac{d(\sfrac{M}{L})}{d\epsilon}$ rather than $\sfrac{d(\sfrac{M_*}{L})}{d\epsilon} \simeq 0.01\sfrac{d(\sfrac{M}{L})}{d\epsilon}$ as observed.
This strengthens the conclusion that the positive (non-null)  $\sfrac{M}{L}$-$\epsilon$ correlation, found in~\cite{Deur:2013baa} and the present analysis, is genuine. At the least, it shows that the correlation is not due to an observational or methodological bias that would also affect $\sfrac{M_*}{L}$. 
 
\section{Summary and conclusion}

We used recent extractions of the dark mass of elliptical galaxies to look for a possible correlation with the galactic ellipticity $\epsilon$. 
We followed the method employed in~\cite{Deur:2013baa} which selects a sample of galaxies as large and homogeneous as possible. 
Uncorrelated systematic effects are suppressed statistically and minimized by the sample homogeneity. 
The dark masses of the 237 galaxies used in this analysis are obtained from 5
different approaches:
(stellar orbit modeling, 
gravitational lensing,
orbits of globular clusters, X-ray emission, and 
orbits of companion galaxies), 
which reduces the possibility of methodological bias.
We found a correlation of $\langle \sfrac{d(\sfrac{M}{L})}{d\epsilon}\rangle = $ $14.1 \pm 5.4 \;\sfrac{M_\odot}{L_\odot}$, which agrees with $14.5\pm4.8 \; \sfrac{M_\odot}{L_\odot}$ from~\cite{Deur:2013baa}.

As a global check for possible effects of observational, measurement, or methodological biases, stellar $\sfrac{M_*}{L}$ ratios were also analyzed and were found to have no significant correlation with $\epsilon$. This suggests
that the $\sfrac{M}{L}$-$\epsilon$ correlation found is genuine.
Several conclusions are possible from such non-zero correlation:

    1. The dark matter halo exerts a strong influence on
a galaxy shape, perhaps because the halo is itself asymmetric.
If so, this would allow us experimental access to the dark halo shape. 

2. There is a significant bias in the current data or methods used to access DMC (lensing, JAM, etc.
and they cannot be trusted to estimate accurately the DMC of elliptical
galaxies. However, the thorough checks performed in~\cite{Deur:2013baa} and here suggest
a genuine correlation rather than an observational or methodological issue.

3. The dynamical evidences from which the DMC of galaxies
is inferred are misinterpreted. In fact, the stimulus for investigating the $\epsilon$-dependence of $\sfrac{M}{L}$
originated from a prediction from Ref.~\cite{Deur:2009ya}. 
In this framework, a homogeneous system locally dense enough so that the non-linearity
of General Relativity is not negligible, should display a correlation between its dynamical total mass
analyzed using Newton's gravity 
and its spacial asymmetry
(e.g., the galaxy ellipticity). Besides the present correlation, this framework also explains~\cite{Deur:2019kqi} the
correlation between dynamical and baryonic matter accelerations observed in Ref.~\cite{McGaugh}, the Cosmic Microwave Background anisotropies, matter power spectrum~\cite{Deur:2022ooc}, large structure formation~\cite{Deur:2021ink}, and unifies the origins of dark matter and dark energy~\cite{Deur:2017aas}.
    

If there is truly a genuine correlation, once the total/dynamical galactic mass is known, the true ellipticity
of the galaxy can be directly obtained. This offers a practical application of our results since the true ellipticity is more delicate to assess comparatively to the DMC.

~

\section*{Acknowledgments}  
This work is supported by the Ingrassia Family Research Grant at the University of Virginia. This research has made use of the NASA/IPAC
Extragalactic Database (NED) which is operated by the Jet Propulsion
Laboratory, California Institute of Technology, under contract with
the National Aeronautics and Space Administration. 
We are grateful to
N. Ouellette, 
S. Phillipps, and 
Cheng-Yu Chen for giving us access to their data, and to
B. Terzi\'c and C. Sargent for their 
useful comments on the manuscript. 



\bibliographystyle{spphys}

\bibliography{References_new}

\section*{Appendix} 
We show here the remaining DMC vs $\epsilon$ from the 13 articles that were not included in Fig.~\ref{fig: MLplots}.

\includegraphics[scale=.35]{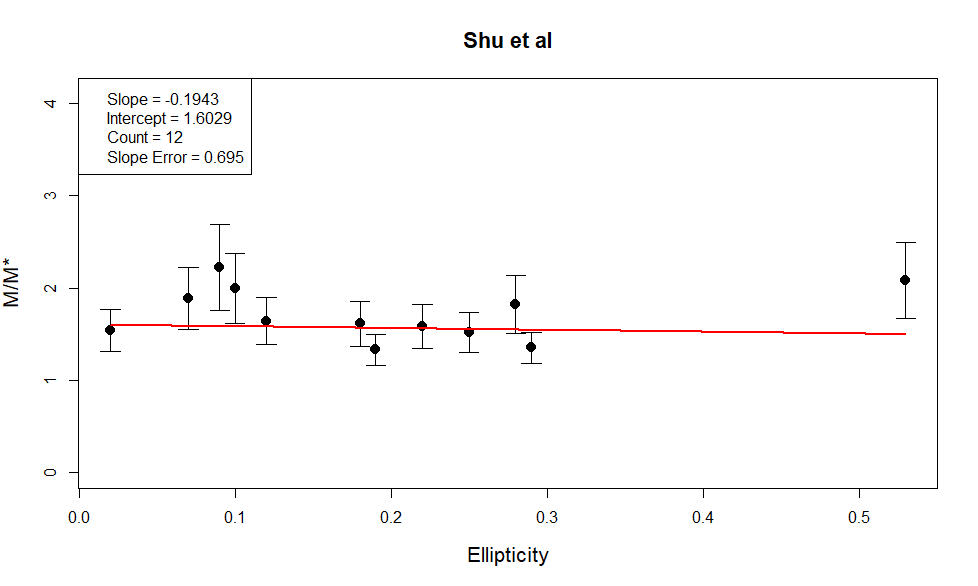}
\includegraphics[scale=.35]{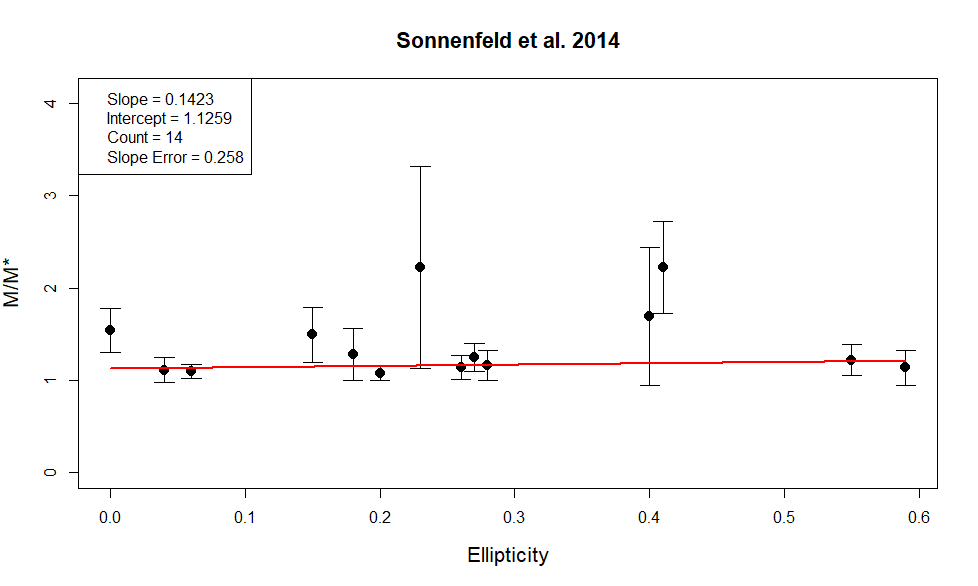}
\includegraphics[scale=.35]{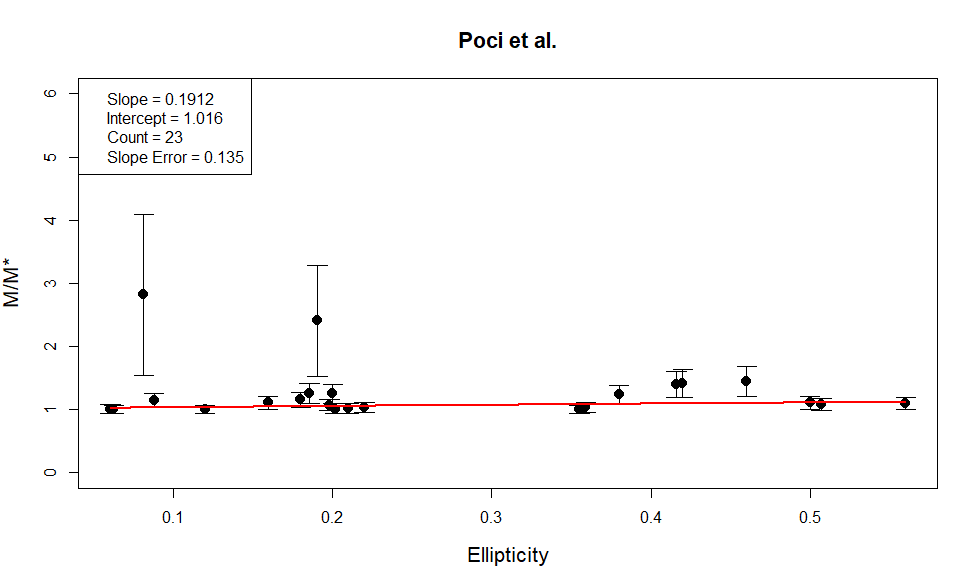}
\includegraphics[scale=.35]{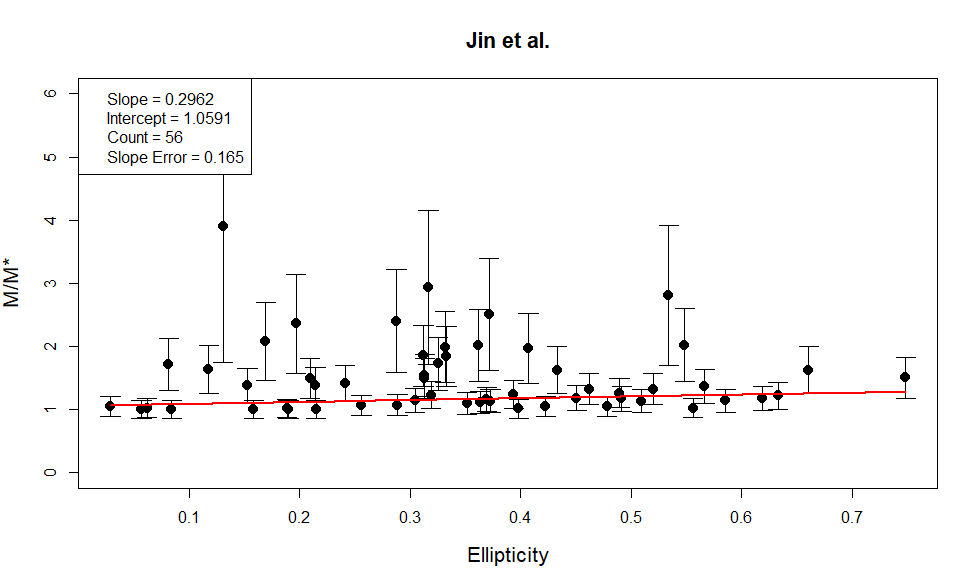}
\includegraphics[scale=.35]{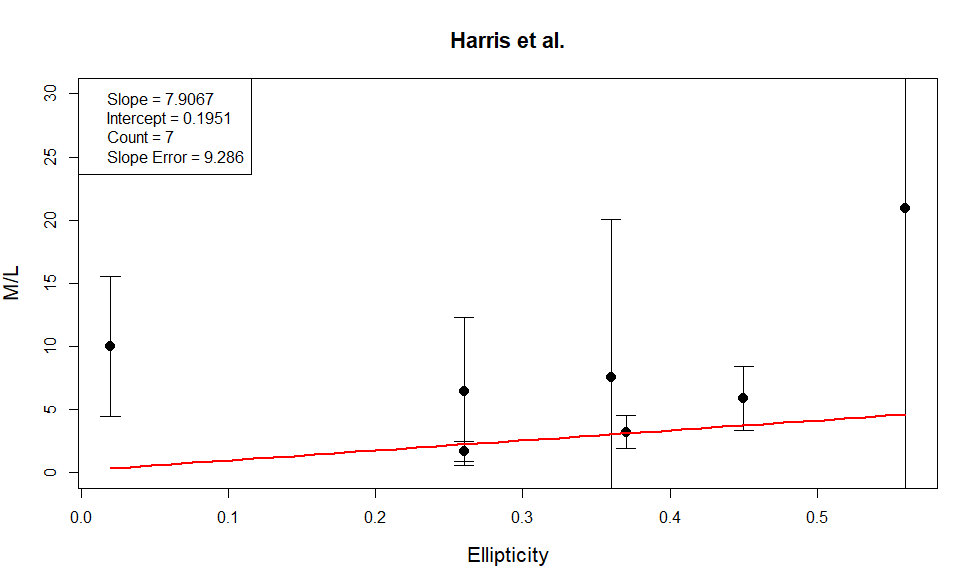}
\includegraphics[scale=.35]{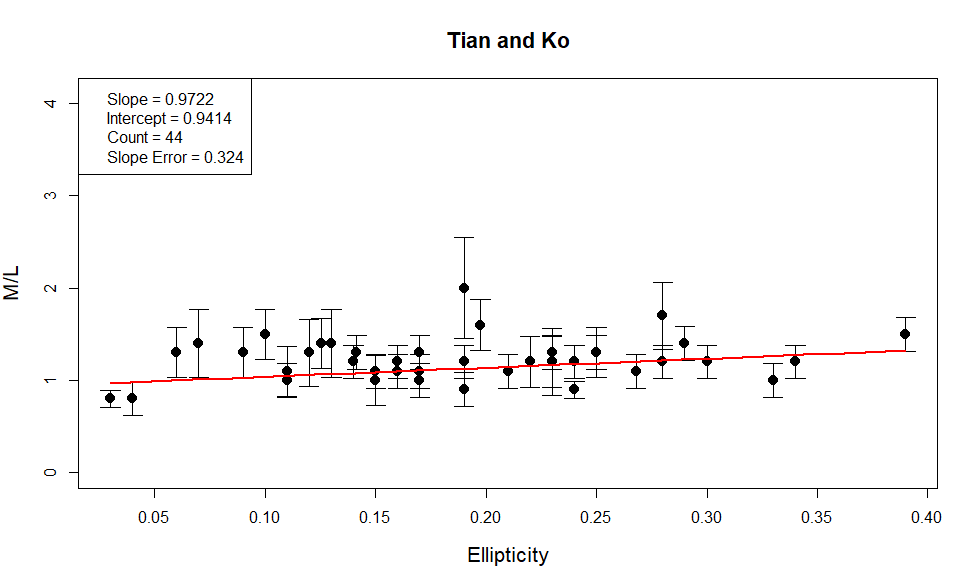}
\includegraphics[scale=.35]{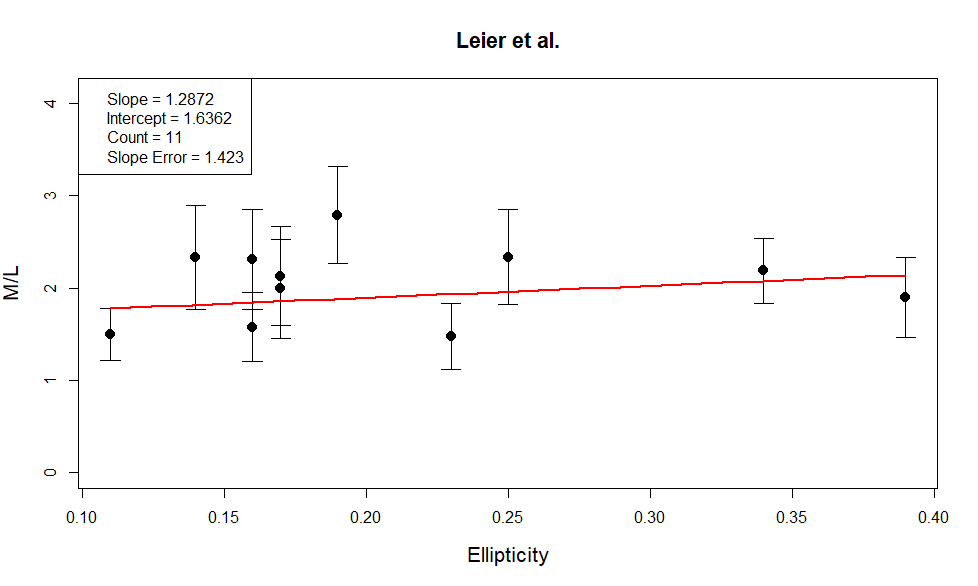}
\includegraphics[scale=.35]{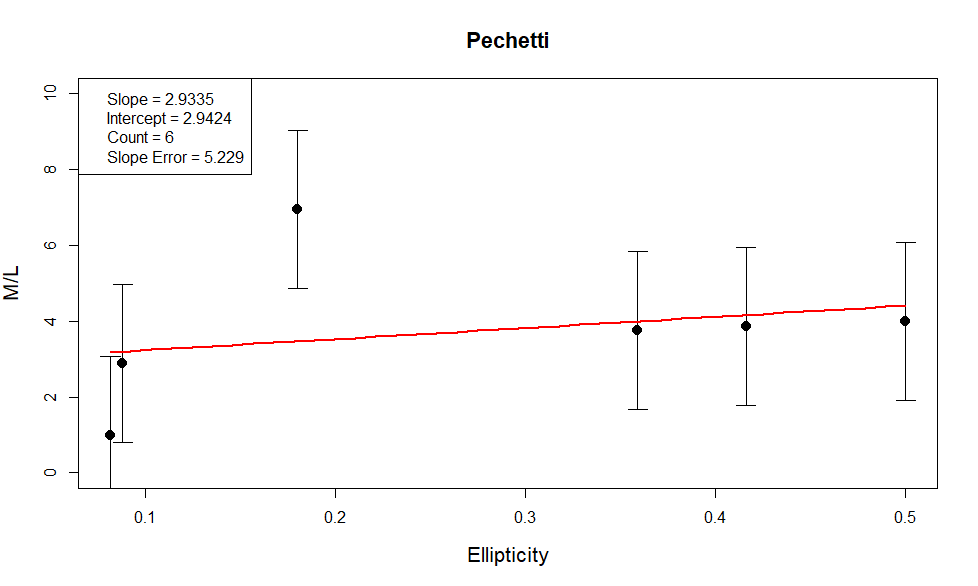}
\includegraphics[scale=.35]{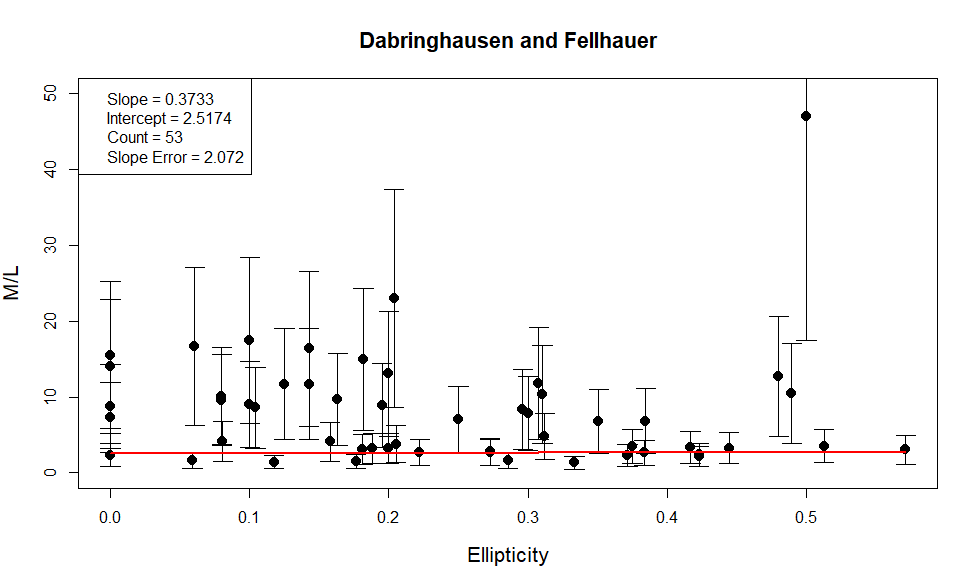}
\includegraphics[scale=.35]{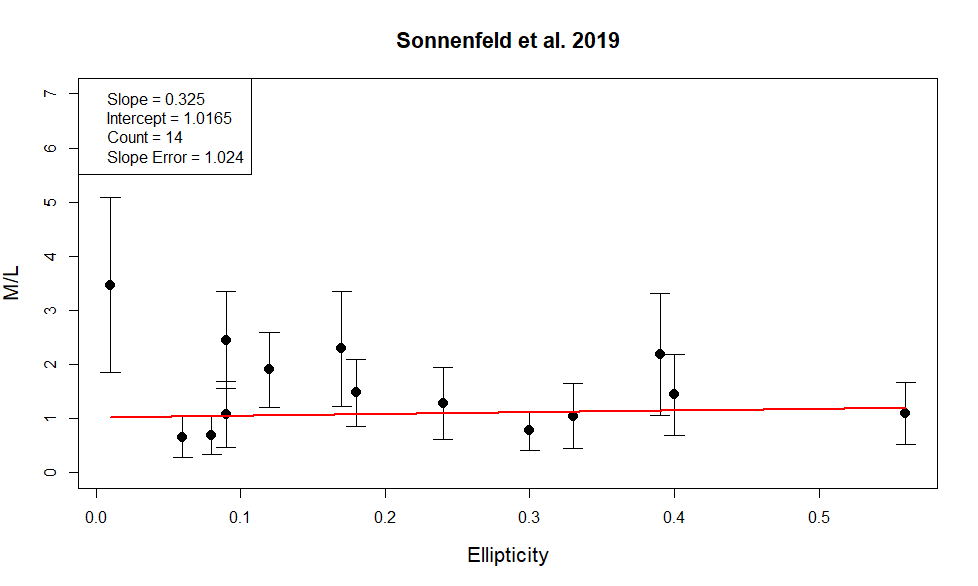}
\includegraphics[scale=.35]{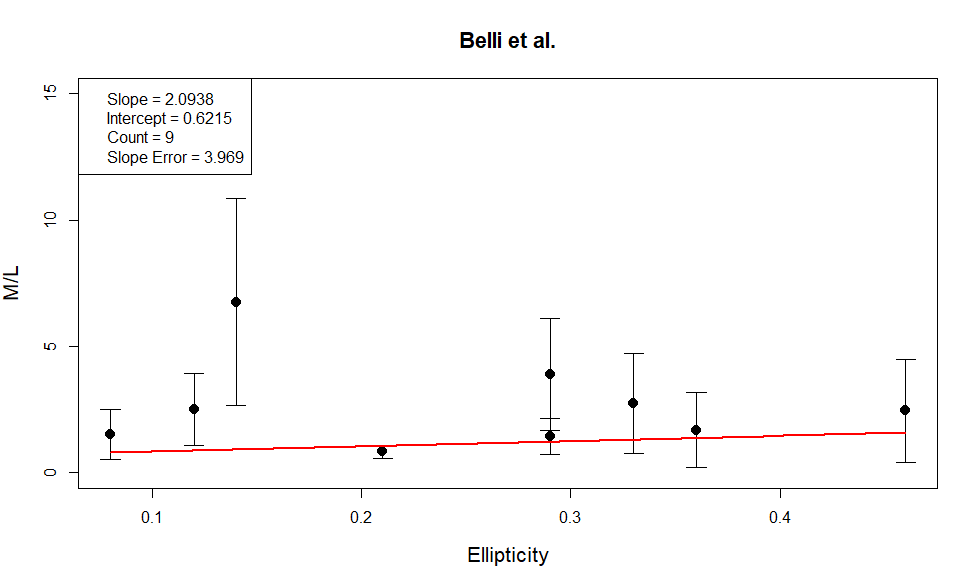}
\includegraphics[scale=.35]{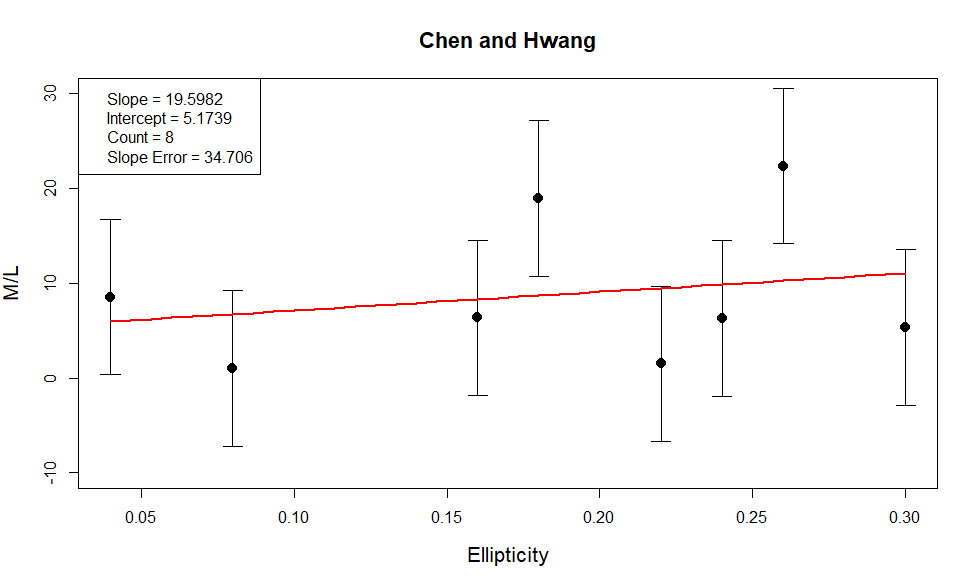}
\includegraphics[scale=.35]{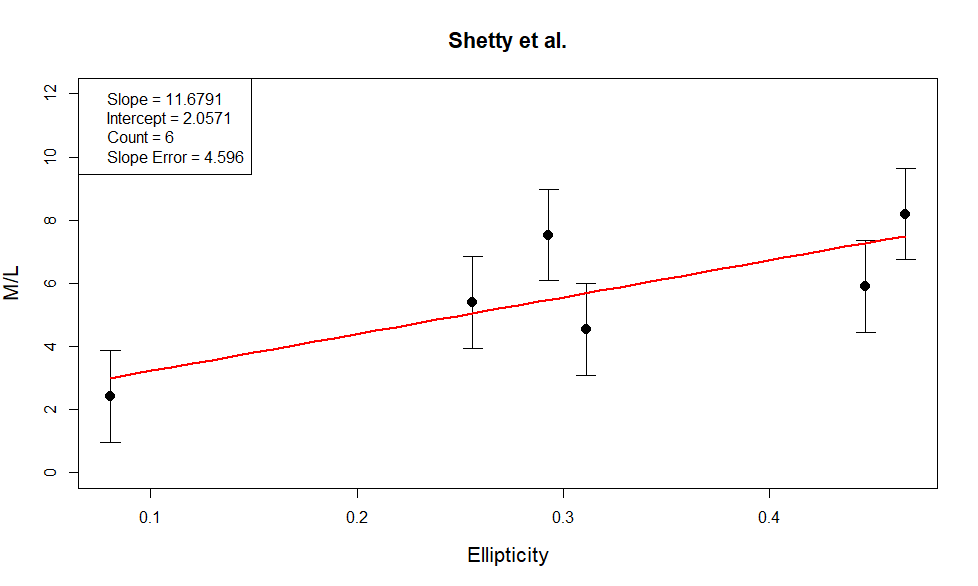}

\end{document}